# New pure shear elastic surface waves in magneto-electro-elastic half-space

Arman Melkumyan [*]

*Department of Mechanics, Yerevan State University, Alex Manoogyan Street 1,*

*Yerevan 375025, Armenia*

**Abstract**

Pure shear surface waves guided by the free surface of a magneto-electro-elastic material are investigated. Three surface waves are obtained for various magneto-electrical conditions on the free surface of the magneto-electro-elastic half-space. The velocities of propagation and the existence conditions for each of these waves are obtained in explicit exact form.

## 1. Introduction

Surface acoustic wave (SAW) devices are widely used in numerous branches of science and technology and their investigation especially in the case of interconnected physical fields in general is an important and actively developing branch of research and applications. About 120 years ago the first type of SAW was described by Lord Rayleigh (1885) in connection with the problem of earthquakes. Bleustein (1968) and Gulyaev (1969) for the first time theoretically predicted that a pure shear SAW can be guided by the free surface of a piezoelectric half-space. Their fundamental results lay in the bases of further developments of acoustoelectronics and currently are cited in more than 350 original papers.

[*] Tel. (+374 91) 482977.
 *E-mail address:* melk_arman@yahoo.com

Recent developments in physics and technology made possible to construct new materials called magneto-electro-elastic materials which demonstrate interconnection between magnetic, electric and elastic fields (Nan, 1994; Srinivas et al. 2006). When the electric field was connected with the elastic one (the piezoelectric materials) it brought up new and unexpected possibilities for science and technology. Connecting the magnetic field with the electric and elastic ones in magneto-electro-elastic materials suggests a range of new possibilities.

In this paper an investigation on the existence of pure shear surface acoustic waves in a magneto-electro-elastic half-space is conduced. It is shown that three different surface acoustic waves can be guided by the free surface of the magneto-electro-elastic half-space in the cases of different magneto-electrical boundary conditions. The first SAW disappears and the other two degenerate to the Bleustein-Gulyaev surface wave when the piezomagnetic and the magnetoelectric coefficients tend to zero, thereby recovering the situation present in piezoelectric materials. These results are important in the further investigation of the dynamic behavior of magneto-electro-elastic materials and will have numerous applications in SAW devices.

**2. General equations**

Let $x_1$, $x_2$, $x_3$ denote rectangular Cartesian coordinates with $x_3$ oriented in the direction of the sixfold axis of a transversely isotropic magneto-electro-elastic material in class 6 mm. Introducing electric potential $\varphi$ and magnetic potential $\phi$, so that

$$\mathbf{E}(x,y,t) = -\nabla \cdot \varphi(x,y,t), \quad \mathbf{H}(x,y,t) = -\nabla \cdot \phi(x,y,t), \tag{1}$$

where $\mathbf{E}$ is the electric field vector and $\mathbf{H}$ is the magnetic field vector, the five partial differential equations which govern the mechanical displacements $u_1$, $u_2$, $u_3$, and the potentials $\varphi$, $\phi$, reduce to two sets of equations when motions independent of the $x_3$ coordinate are considered. The equations of interest in the present paper are those governing the $u_3$ component of the displacement and the potentials $\varphi$, $\phi$, and can be written in the following form (Li, 2005):

$$c_{44}\nabla^2 u_3 + e_{15}\nabla^2 \varphi + q_{15}\nabla^2 \phi = \rho \ddot{u}_3 ,$$
$$e_{15}\nabla^2 u_3 - \varepsilon_{11}\nabla^2 \varphi - d_{11}\nabla^2 \phi = 0 , \tag{2}$$

$$q_{15}\nabla^2 u_3 - d_{11}\nabla^2\varphi - \mu_{11}\nabla^2\phi = 0 ,$$

where $\nabla^2$ is the two-dimensional Laplacian operator, $\nabla^2 = \partial^2/\partial x_1^2 + \partial^2/\partial x_2^2$, $\rho$ is the mass density, $c_{44}$, $e_{15}$, $\varepsilon_{11}$, $q_{15}$, $d_{11}$ and $\mu_{11}$ are respectively the elastic modulus, piezoelectric coefficient, dielectric constant, piezomagnetic coefficient, magnetoelectric coefficient and magnetic permittivity constant, and the superposed dot indicates differentiation with respect to time. The constitutive equations which relate the stresses $T_{ij}$ ($i, j = 1, 2, 3$), the electric displacements $D_i$ ($i = 1, 2, 3$) and the magnetic induction $B_i$ ($i = 1, 2, 3$) to $u_3$, $\varphi$ and $\phi$ are

$$T_1 = T_2 = T_3 = T_{12} = 0, \quad D_3 = 0, \quad B_3 = 0 ,$$

$$T_{23} = c_{44}u_{3,2} + e_{15}\varphi_{,2} + q_{15}\phi_{,2} , \quad T_{13} = c_{44}u_{3,1} + e_{15}\varphi_{,1} + q_{15}\phi_{,1} ,$$

$$D_1 = e_{15}u_{3,1} - \varepsilon_{11}\varphi_{,1} - d_{11}\phi_{,1} , \quad D_2 = e_{15}u_{3,2} - \varepsilon_{11}\varphi_{,2} - d_{11}\phi_{,2} , \qquad (3)$$

$$B_1 = q_{15}u_{3,1} - d_{11}\varphi_{,1} - \mu_{11}\phi_{,1} , \quad B_2 = q_{15}u_{3,2} - d_{11}\varphi_{,2} - \mu_{11}\phi_{,2} .$$

Solving Eqs. (2) for $\nabla^2 u_3$, $\nabla^2\varphi$ and $\nabla^2\phi$ and defining functions $\psi$ and $\chi$ by

$$\psi = \varphi - m u_3 , \quad \chi = \phi - n u_3 , \qquad (4)$$

the solution of Eqs. (2) is reduced to the solution of

$$\nabla^2 u_3 = \rho \tilde{c}_{44}^{-1} \ddot{u}_3 , \quad \nabla^2 \psi = 0 , \quad \nabla^2 \chi = 0 , \qquad (5)$$

where

$$m = \frac{e_{15}\mu_{11} - q_{15}d_{11}}{\varepsilon_{11}\mu_{11} - d_{11}^2} , \quad n = \frac{q_{15}\varepsilon_{11} - e_{15}d_{11}}{\varepsilon_{11}\mu_{11} - d_{11}^2} , \qquad (6)$$

and

$$\tilde{c}_{44} = c_{44} + \frac{e_{15}^2\mu_{11} - 2e_{15}q_{15}d_{11} + q_{15}^2\varepsilon_{11}}{\varepsilon_{11}\mu_{11} - d_{11}^2}$$

$$= \bar{c}_{44}^e + \frac{(d_{11}e_{15} - q_{15}\varepsilon_{11})^2}{\varepsilon_{11}(\varepsilon_{11}\mu_{11} - d_{11}^2)}$$

$$= \bar{c}_{44}^m + \frac{(d_{11}q_{15} - e_{15}\mu_{11})^2}{\mu_{11}(\varepsilon_{11}\mu_{11} - d_{11}^2)} . \qquad (7)$$

In Eqs. (5), (7) $\tilde{c}_{44}$ is the magneto-electro-elastically stiffened elastic constant, $\bar{c}_{44}^e = c_{44} + e_{15}^2/\varepsilon_{11}$ is the electro-elastically stiffened elastic constant and $\bar{c}_{44}^m = c_{44} + q_{15}^2/\mu_{11}$ is the magneto-elastically stiffened elastic constant.

Together with the electro-mechanical coupling coefficient $k_e^2 = e_{15}^2 / (\varepsilon_{11} \bar{c}_{44}^e)$ and the magneto-mechanical coupling coefficient $k_m^2 = q_{15}^2 / (\mu_{11} \bar{c}_{44}^m)$ introduce the magneto-electro-mechanical coupling coefficient

$$k_{em}^2 = \frac{e_{15}^2 \mu_{11} - 2 e_{15} q_{15} d_{11} + q_{15}^2 \varepsilon_{11}}{\tilde{c}_{44} \left( \varepsilon_{11} \mu_{11} - d_{11}^2 \right)}$$

$$= \frac{e_{15}^2}{\tilde{c}_{44} \varepsilon_{11}} + \frac{(q_{15} \varepsilon_{11} - e_{15} d_{11})^2}{\tilde{c}_{44} \varepsilon_{11} \left( \varepsilon_{11} \mu_{11} - d_{11}^2 \right)}$$

$$= \frac{q_{15}^2}{\tilde{c}_{44} \mu_{11}} + \frac{(e_{15} \mu_{11} - q_{15} d_{11})^2}{\tilde{c}_{44} \mu_{11} \left( \varepsilon_{11} \mu_{11} - d_{11}^2 \right)} . \tag{8}$$

From Eqs. (6)-(8) it follows that

$$\tilde{c}_{44} = c_{44} / \left(1 - k_{em}^2\right),$$

$$(e_{15} \mu_{11} - q_{15} d_{11}) m = \tilde{c}_{44} \mu_{11} k_{em}^2 - q_{15}^2 ,$$

$$(q_{15} \varepsilon_{11} - e_{15} d_{11}) n = \tilde{c}_{44} \varepsilon_{11} k_{em}^2 - e_{15}^2 , \tag{9}$$

$$e_{15} m + q_{15} n = \tilde{c}_{44} k_{em}^2 , \quad \varepsilon_{11} m + d_{11} n = e_{15} , \quad d_{11} m + \mu_{11} n = q_{15} .$$

Using the introduced functions $\psi$ and $\chi$ and the magneto-electro-elastically stiffened elastic constant, the constitutive Eqs. (3) can be written in the following form:

$$T_1 = T_2 = T_3 = T_{12} = 0 , \; D_3 = 0 , \; B_3 = 0 ,$$

$$T_{23} = \tilde{c}_{44} u_{3,2} + e_{15} \psi_{,2} + q_{15} \chi_{,2} , \quad T_{13} = \tilde{c}_{44} u_{3,1} + e_{15} \psi_{,1} + q_{15} \chi_{,1} ,$$

$$D_1 = -\varepsilon_{11} \psi_{,1} - d_{11} \chi_{,1} , \quad D_2 = -\varepsilon_{11} \psi_{,2} - d_{11} \chi_{,2} , \tag{10}$$

$$B_1 = -d_{11} \psi_{,1} - \mu_{11} \chi_{,1} , \quad B_2 = -d_{11} \psi_{,2} - \mu_{11} \chi_{,2} .$$

Using Eqs. (3) and the condition of the positiveness of energy one has that

$$c_{44} > 0, \; \varepsilon_{11} > 0, \; \mu_{11} > 0, \; \varepsilon_{11} \mu_{11} - d_{11}^2 > 0. \tag{11}$$

From Eqs. (7), (8) and (11) one has that

$$\tilde{c}_{44} \geq c_{44}, \; \tilde{c}_{44} \geq \bar{c}_{44}^e, \; \tilde{c}_{44} \geq \bar{c}_{44}^m;$$

$$\tilde{c}_{44} = c_{44} \text{ if and only if } k_{em} = 0;$$

$$\tilde{c}_{44} = \bar{c}_{44}^e \text{ if and only if } n = 0; \tag{12}$$

$$\tilde{c}_{44} = \bar{c}_{44}^m \text{ if and only if } m = 0;$$

and

$$k_{em}^2 \geq \frac{e_{15}^2}{\tilde{c}_{44}\varepsilon_{11}}, \quad k_{em}^2 \geq \frac{q_{15}^2}{\tilde{c}_{44}\mu_{11}}, \quad 0 \leq k_{em} < 1;$$

$$k_{em}^2 = \frac{e_{15}^2}{\tilde{c}_{44}\varepsilon_{11}} \text{ if and only if } n = 0;$$

$$k_{em}^2 = \frac{q_{15}^2}{\tilde{c}_{44}\mu_{11}} \text{ if and only if } m = 0; \qquad (13)$$

$$k_{em} = 0 \text{ if and only if } e_{15} = 0, \quad q_{15} = 0.$$

If $d_{11} = 0$ then the expressions of the magneto-electro-elastically stiffened elastic constant and the magneto-electro-mechanical coupling coefficient are simplified:

$$\tilde{c}_{44} = c_{44} + \frac{e_{15}^2}{\varepsilon_{11}} + \frac{q_{15}^2}{\mu_{11}} = \bar{c}_{44}^e + \bar{c}_{44}^m - c_{44},$$

$$k_{em}^2 = \frac{e_{15}^2}{\tilde{c}_{44}\varepsilon_{11}} + \frac{q_{15}^2}{\tilde{c}_{44}\mu_{11}}. \qquad (14)$$

## 3. Surface waves

The mechanical condition for the free surface of the magneto-electro-elastic material which occupies the half-space $x_2 \geq 0$ is

$$T_{23} = \tilde{c}_{44}u_{3,2} + e_{15}\psi_{,2} + q_{15}\chi_{,2} = 0 \text{ on } x_2 = 0, \qquad (15)$$

which must be satisfied together with the magneto-electrical conditions of electrically closed surface ($\varphi = 0$), electrically open surface ($D_2 = 0$), magnetically closed surface ($B_2 = 0$) or magnetically open surface ($\phi = 0$).

First we study the case of the mechanically free, electrically and magnetically open surface $x_2 = 0$ of the half-space $x_2 \geq 0$. The corresponding boundary conditions on $x_2 = 0$ are

$$T_{23} = \tilde{c}_{44}u_{3,2} + e_{15}\psi_{,2} + q_{15}\chi_{,2} = 0,$$

$$D_2 = -\varepsilon_{11}\psi_{,2} - d_{11}\chi_{,2} = 0, \qquad (16)$$

$$\phi = \chi + nu_3 = 0.$$

The conditions at infinity require that

$$u_3, \varphi, \phi \to 0 \text{ as } x_2 \to \infty. \tag{17}$$

Consider the possibility of a solution of Eqs. (5) of the form:

$$u_3 = A\exp(-\xi_2 x_2)\exp[i(\xi_1 x_1 - \omega t)],$$

$$\psi = B\exp(-\xi_1 x_2)\exp[i(\xi_1 x_1 - \omega t)], \tag{18}$$

$$\chi = C\exp(-\xi_1 x_2)\exp[i(\xi_1 x_1 - \omega t)],$$

which represents a SAW propagating in the positive direction of the $x_1$ axis.

These expressions identically satisfy the second and the third of Eqs. (5) and the first of Eqs. (5) requires

$$\tilde{c}_{44}\left(\xi_1^2 - \xi_2^2\right) = \rho\omega^2. \tag{19}$$

The conditions at infinity (17) are satisfied if $\xi_1 > 0$ and $\xi_2 > 0$.

The boundary conditions in Eqs. (16) require

$$A(\tilde{c}_{44}\xi_2) + B(e_{15}\xi_1) + C(q_{15}\xi_1) = 0,$$

$$B(\varepsilon_{11}\xi_1) + C(d_{11}\xi_1) = 0, \tag{20}$$

$$A\frac{q_{15}\varepsilon_{11} - e_{15}d_{11}}{\varepsilon_{11}\mu_{11} - d_{11}^2} + C = 0.$$

For nontrivial $A$, $B$, $C$ the determinant of the coefficients of $A$, $B$ and $C$ in Eqs. (20) must be equal to zero which leads to the relation

$$\xi_2 = \left(k_{em}^2 - \frac{e_{15}^2}{\varepsilon_{11}\tilde{c}_{44}}\right)\xi_1, \tag{21}$$

so that using Eq. (19) the velocity $V_{s1} = \omega/\xi_1$ of propagation of the first surface wave is determined:

$$V_{s1}^2 = \frac{\tilde{c}_{44}}{\rho}\left(1 - \left[k_{em}^2 - \frac{e_{15}^2}{\varepsilon_{11}\tilde{c}_{44}}\right]^2\right). \tag{22}$$

These results show that a pure shear surface wave can exist in a mechanically free and electrically open magneto-electro-elastic half-space, whereas there is no surface wave with $u_3 \neq 0$, $u_1 = u_2 = 0$ in a half-space with such boundary conditions in piezoelectric materials. This is easily demonstrated by setting $d_{11} = 0$, $q_{15} = 0$ in Eq.

(21), from which we conclude that $\xi_2 = 0$ and hence the displacement amplitude does not decay as we move away from the surface (a degenerate member of the family of SH waves).

In the case of the mechanically free, electrically and magnetically closed surface $x_2 = 0$ the nontrivial boundary conditions on $x_2 = 0$ are:

$$T_{23} = \tilde{c}_{44}u_{3,2} + e_{15}\psi_{,2} + q_{15}\chi_{,2} = 0,$$

$$\varphi = \psi + mu_3 = 0, \tag{23}$$

$$B_2 = -d_{11}\psi_{,2} - \mu_{11}\chi_{,2} = 0.$$

Proceeding in the same manner as in the previous problem we show that a nontrivial solution of the form (18) is possible if Eq. (19) is satisfied and $\xi_2 = \left[k_{em}^2 - q_{15}^2/(\mu_{11}\tilde{c}_{44})\right]\xi_1$, so that the second surface wave velocity of the magneto-electro-elastic half-space is determined:

$$V_{s2}^2 = \frac{\tilde{c}_{44}}{\rho}\left(1 - \left[k_{em}^2 - \frac{q_{15}^2}{\mu_{11}\tilde{c}_{44}}\right]^2\right). \tag{24}$$

In the case of the mechanically free, electrically closed and magnetically open surface $x_2 = 0$ the nontrivial boundary conditions on $x_2 = 0$ are:

$$T_{23} = \tilde{c}_{44}u_{3,2} + e_{15}\psi_{,2} + q_{15}\chi_{,2} = 0,$$

$$\varphi = \psi + mu_3 = 0, \tag{25}$$

$$\phi = \chi + nu_3 = 0,$$

which lead to the condition $\xi_2 = k_{em}^2\xi_1$ and the third surface wave velocity is found to be

$$V_{s3}^2 = \frac{\tilde{c}_{44}}{\rho}\left(1 - k_{em}^4\right). \tag{26}$$

In the case of the mechanically free, electrically open and magnetically closed surface $x_2 = 0$ the corresponding analysis shows that no surface wave can be guided by the surface of the magneto-electro-elastic half-space.

From Eqs. (13), (17), (19), (22), (24), (26) follows that the existence criteria for the three surface waves are the following:

$n \neq 0$ for existence of the surface wave with velocity $V_{s1}$;

$m \neq 0$ for existence of the surface wave with velocity $V_{s2}$; (27)

$k_{em} \neq 0$ for existence of the surface wave with velocity $V_{s3}$.

The surface wave with velocity $V_{s1}$ disappears and the surface wave velocities $V_{s2}$ and $V_{s3}$ which are different in the case of magneto-electro-elastic materials tend to the same value of $V_{bg} = \sqrt{(\bar{c}_{44}^e/\rho)(1-k_e^4)}$ i.e. the Bleustein-Gulyaev surface wave's speed when $d_{11}$ and $q_{15}$ tend to zero. It shows that these two different surface waves of the magneto-electro-elastic materials merge with each other and recover the Bleustein-Gulyaev surface wave present in the piezoelectric materials when the electromagnetic constant $d_{11}$ and the piezomagnetic constant $q_{15}$ of the magneto-electro-elastic materials vanish.

## 4. Conclusions

The existence of pure shear surface waves in magneto-electro-elastic materials is investigated. It is shown that three surface waves can be guided by the free surface of a transversely isotropic magneto-electro-elastic half-space in the cases of different magneto-electrical boundary conditions. The existence criteria and the propagation velocities of each of these three waves are obtained in explicit exact forms. The first surface wave vanishes, while the other two surface waves merge and recover the Bleustein-Gulyaev surface wave of piezoelectric materials when the piezomagnetic and the magnetoelectric coefficients tend to zero.